\documentclass[review,12pt]{elsarticle}

\usepackage{amssymb}
\usepackage{amsthm,amsmath}
\usepackage{accents}
\usepackage{graphicx}
\usepackage{epstopdf}
\newtheorem{theorem}{Theorem}

\newtheorem{Definition}{Definition}

\newtheorem{propo}{Proposition}

\begin{document}
\begin{center}
{\Large  Finite time Convergence of Pinning Synchronization with a single Nonlinear Controller}

\footnote{This work is jointly supported by the National Key R$\&$D Program of China (No.
2018AAA010030), National Natural Sciences Foundation of China under Grant (No.
61673119 and 61673298), STCSM (No. 19JC1420101), Shanghai Municipal Science and
Technology Major Project under Grant 2018SHZDZX01 and ZJLab, the Key Project of
Shanghai Science and Techonology under Grant 16JC1420402.9).
}
\\[0.2in]
\begin{center}
Tianping Chen\footnote{Tianping Chen is with the School of Computer
Sciences/Mathematics, Fudan University, 200433, Shanghai, China. \\
\indent ~~Corresponding author: Tianping Chen. Email:
tchen@fudan.edu.cn}, Wenlian Lu \footnote{
Wenlian Lu is with the Centre for Computational
Systems Biology and School of Mathematical Sciences, Fudan University, People's Republic of China (wenlian@fudan.edu.cn)},
Xiwei Liu \footnote{Xiwei Liu is with Department of Computer Science and
Technology, Tongji University, and with the Key Laboratory of Embedded System and Service Computing,
Ministry of Education, Shanghai 200092, China. E-mail:
xwliu@tongji.edu.cn},

\end{center}
\end{center}

\begin{abstract}
In this paper, we discuss distributive synchronization of complex networks in finite time, with  a single nonlinear pinning controller. The results apply to heterogeneous dynamic networks, too. Different from many models, which assume the coupling matrix being symmetric (or the connecting graph is undirected), here, the coupling matrix is asymmetric (or the connecting graph is directed).

\end{abstract}


\section{Introduction}
Synchronization of complex networks has been a hot topic in recent decades. The model in the  literature can be described as
\begin{align}
\dot{x}_i(t)=f(x_i(t))+c\sum_{j=1}^{m} a_{ij}x_j(t)
\end{align}
where $x_{i}(t)=[x_{i}^{1}(t),\cdots,x_{i}^{n}(t)]^{T}\in R^{n}$, $i=1,\cdots m$.

The purpose of synchronization is to make $x_{i}(t)-x_{j}(t)\rightarrow 0$ for all $x_{i}(t)$, $i=1,\cdots,m$, when $t\rightarrow +\infty$.


In some cases, one hopes to make all $x_{i}(t)-s(t)\rightarrow 0$ when $t\rightarrow +\infty$, $i=1,\cdots,m$, and $s(t)$ is a solution of
\begin{align}\label{target}
\dot{s}(t)=f(s(t)).
\end{align}

Pick $\varepsilon>0$, in \cite{Chen1}, the following coupled network
with a single linear controller
\begin{eqnarray}\label{pina}
\left\{\begin{array}{cl}
\frac{dx_1(t)}{dt}&=f(x_1(t))
+c\sum\limits_{j=1}^ma_{1j}x_j(t)-c\varepsilon(x_{1}(t)-s(t)),\\
\frac{dx_i(t)}{dt}&=f(x_i(t))+c\sum\limits_{j=1}^ma_{ij}x_j(t), ~~~~i=2,\cdots,m \end{array}\right.
\end{eqnarray}
was proposed. It can make all $x_{i}(t)-s(t)\rightarrow 0$ exponentially, if $c$ is chosen suitably.


Another topic is finite time convergence \cite{Finite86}-\cite{Finite16}. Finite time consensus was studied in \cite{Finite06} and \cite{Finite09}, where the general protocol has the following form:
\begin{align*}
\dot{x}_i(t)=\beta sig(\sum_{j}a_{ij}x_j(t))^{\alpha}+\gamma\sum_ja_{ij}x_j(t)
\end{align*}
with $0\le \alpha<1$, $\beta>0$, and $\gamma\ge 0$, and the network is directed.

On the other hand, \cite{Finite13} added controllers on all the nonidentical nodes and realized the finite time synchronization; \cite{Finite18} considered the finite/fixed time cluster synchronization for undirected networks with or without control; \cite{Finite19} considered the exponential synchronization for switched networks with pinning technique; \cite{Finite20} studied the finite time cluster synchronization in complex-variable, fractional-order and undirected networks; \cite{Finite20a} and \cite{Finite21} investigated the finite time synchronization/stabilization for the (drive-response) inertial memristive neural networks with mixed delays. In these models, nonlinear couplings or controllers are introduced. However, the coupling matrices are required symmetric and/or the nonlinear controllers are added to all nodes.

It is natural to raise a question:

Can we design new algorithms with only a single pinning controller to make all $x_{i}(t)-s(t)\rightarrow 0$ in finite time for directly connected networks?

In this paper, we give a confirmative answer. We propose simple finite time synchronization models and prove that under mild conditions, they can reach finite time synchronization.

\section{Some basic concepts and background}

Denote $x_{i}=[x_{i}^{1},\cdots,x_{i}^{n}]^{T}\in R^{n}$, $i=1,\cdots,m$. $x=[x_{1}^{T},\cdots,x_{m}^{T}]^{T}\in R^{m n}$. $||x_{i}||_{1}=\sum_{k=1}^{n}|x_{i}^{k}|$. $||x||_{1}=\sum_{i=1}^{m}||x_{i}||_{1}=\sum_{i=1}^{m}\sum_{k=1}^{n}|x_{i}^{k}|$. 

\begin{Definition} A coupling matrix $A=(a_{ij})_{i,j=1}^{m}$ is defined as a Metzler matrix satisfying $a_{ij}\ge 0$, if $i\ne j$ and $a_{ii}=-\sum_{j\ne i}a_{ij}$. We also assume $Rank(A)=m-1$.
\end{Definition}
\begin{Definition}  A coupling matrix with a single pinning control on the first node $\tilde{A}=(\tilde{a}_{ij})_{i,j=1}^{m}$ is defined as follows
$\tilde{a}_{11}=a_{11}-\varepsilon$, $\varepsilon>0$ and
$\tilde{a}_{ij}=a_{ij}$ otherwise.
\end{Definition}

It was revealed in \cite{Chen1}, $-\tilde{A}$ is an M-matrix. By M-matrix theory, there are constants $\theta_{i}>0$, $i=1,\cdots,m$, with $\sum_{i=1}^{m}\theta_{i}=1$ and $\theta^{\star}>0$,  such that 
\begin{align}\label{M1}
\sum_{i=1}^{m}\theta_{i}\tilde{a}_{ij}<-\theta^{\star},
\end{align}
for all $j=1,\cdots,m$.

\section{Distributed finite time synchronization algorithms of complex networks with a single nonlinear controller}

In this section, we introduce distributed discontinuous and continuous algorithms for finite time synchronization of complex networks with a single nonlinear controller.

{\bf Firstly}, we propose following distributed algorithm with a single nonlinear controller
\begin{eqnarray}\label{pinf}
\left\{\begin{array}{ll}\frac{dx_1(t)}{dt}&=f(x_1(t))
+c\frac{\sum\limits_{j=1}^m a_{1j}(x_j(t)-s(t))-\epsilon(x_{1}(t)-s(t))}
{||\sum\limits_{j=1}^m a_{1j}(x_j(t)-s(t))-\epsilon(x_{1}(t)-s(t))||_{1}},\\
\frac{dx_i(t)}{dt}&=f(x_i(t))+c\frac{\sum\limits_{j=1}^m a_{ij}x_j(t)}
{||\sum\limits_{j=1}^ma_{ij}x_j(t)||_{1}},~~~~i=2,\cdots,m \end{array}\right.
\end{eqnarray}
and prove the following Theorem.
\begin{theorem} Suppose $||f(x_i(t))-f(s(t))||_{1}\le L||x_i(t)-s(t)||_{1}$, $L>0$, $\forall i$. Then, the controlled system (\ref{pinf})
can synchronize all $x_{i}(t)$ to $s(t)$ in finite time with sufficiently large $c$.
\end{theorem}
{\bf Proof}~~
It is clear that system (\ref{pinf}) can be written as
\begin{eqnarray}\label{pinf1}
\frac{dx_i(t)}{dt}&=f(x_i(t))+c\frac{\sum\limits_{j=1}^m\tilde{a}_{ij}(x_j(t)-s(t))}
{||\sum\limits_{j=1}^m\tilde{a}_{ij}(x_j(t)-s(t))||_{1}},~~i=1,2,\cdots,m
\end{eqnarray}

Denote $y_{i}(t)=\sum\limits_{j=1}^{m}\tilde{a}_{ij} x_{j}(t)$, $\delta{y}_{i}(t)=\sum\limits_{j=1}^{m}\tilde{a}_{ij} (x_{j}(t)-s(t))$. Then, we have
\begin{align}
\dot{\delta}y_{i}(t)=&\sum\limits_{j=1}^{m}\tilde{a}_{ij}\delta f(x_{j}(t))
+c\sum\limits_{j=1}^{m}\tilde{a}_{ij} \frac{\delta{y}_{j}(t)}{||\delta{y}_{j}(t)||_{1}},
\end{align}
where $\delta f(x_{j}(t))=f(x_j(t))-f(s(t))$. Define a norm
\begin{eqnarray*}
||\delta{y}(t)||_{\{\theta,1\}}=\sum_{i=1}^{m}\theta_{i}
||\delta{y}_{i}(t)||_{1}=\sum_{i=1}^{m}\theta_{i}
\sum_{k=1}^{n}|\delta{y}_{i}^{k}(t)|,
\end{eqnarray*}
where $\delta{y}(t)=[\delta{y}_{1}^T(t),\cdots,\delta{y}_{m}^T(t)]^{T}$, $\delta{y}_{i}(t)=[\delta{y}_{i}^{1}(t),\cdots,\delta{y}_{i}^{n}(t)]^T$,
$i=1,\cdots,m$.

Because $(\tilde{a}_{ij})$ is nonsingular, there are constants
$\kappa_{1}$ and $\kappa_{2}$ such that
$$\kappa_{1}||\delta{y}(t)||_{\{\theta,1\}}\le ||\delta{x}(t)||_{\{1\}}\le \kappa_{2}||\delta{y}(t)||_{\{\theta,1\}}$$
where $||\delta{x}(t)||_{\{1\}}=\sum_{i=1}^{m}
||\delta{x}_{i}(t)||_{1}=\sum_{i=1}^{m}
||{x}_{i}(t)-s(t)||_{1}$.

Combining $||f(x)-f(y)||_{1}\le L||x-y||_{1}$, it is easy to see that
\begin{align*}
\sum_{i=1}^{m}\theta_{i}|\sum\limits_{j=1}^{m}\tilde{a}_{ij}\delta f(x_{j}(t))|&\le L\sum_{i=1}^{m}\theta_{i}\sum\limits_{j=1}^{m}|\tilde{a}_{ij}|||\delta x_{j}(t)||_{1}\\
&\le L\max_{i,j}\{|\tilde{a}_{ij}|\}\sum_{j=1}^{m}
||\delta x_{j}(t)||_{1}\\
&= L\max_{i}\{|\tilde{a}_{ii}|\}||\delta{x}(t)||_{\{1\}}\\
&\le L\kappa_{2}\max_{i}\{|\tilde{a}_{ii}|\}||\delta{y}(t)||_{\{\theta,1\}}
\end{align*}
and
\begin{align*}
&\frac{d
||\delta{y}(t)||_{\{\theta,1\}}}{dt}=\sum\limits_{i=1}^{m}
\theta_{i}\frac{d
||\delta{y_{i}}(t)||_{1}}{dt}=\sum\limits_{i=1}^{m}\sum\limits_{k=1}^{n}
\theta_{i}\frac{d
|\delta{y_{i}^{k}}(t)|}{dt}\\
&=\sum\limits_{i=1}^{m}\sum\limits_{k=1}^{n}
\theta_{i}sign(\delta{y_{i}^{k}}(t))\frac{
d\delta{y_{i}^{k}}(t)}{dt}\\&
\le L\kappa_{2}\max_{i}\{|\tilde{a}_{ii}|\}||\delta{y}(t)||_{\{\theta,1\}}\\
&+
c\sum\limits_{i=1}^{m}\theta_{i}\sum\limits_{k=1}^{n}\bigg(\tilde{a}_{ii}sign(\delta{y_{i}^{k}}(t))
\frac{\delta{y}_{i}^{k}(t)}{||\delta{y}_{i}(t)||_{1}}
+\sum\limits_{j=1,j\ne i}^{m}\tilde{a}_{ij}\frac{sign(\delta{y_{i}^{k}}(t))\delta{y}_{j}^{k}(t)}{||\delta{y}_{j}(t)||_{1}}\bigg)\\&
\le L\kappa_{2}\max_{i}\{|\tilde{a}_{ii}|\}||\delta{y}(t)||_{\{\theta,1\}}\\
&+
c\sum\limits_{i=1}^{m}\theta_{i}\sum\limits_{j=1}^{m}\sum\limits_{k=1}^{n}\bigg(\tilde{a}_{ii}
\frac{|\delta{y}_{i}^{k}(t)|}{||\delta{y}_{i}(t)||_{1}}
+\tilde{a}_{ij}
\frac{|\delta{y}_{j}^{k}(t)|}{||\delta{y}_{j}(t)||_{1}}\bigg)
\\&
= L\kappa_{2}\max_{i}\{|\tilde{a}_{ii}|\}||\delta{y}(t)||_{\{\theta,1\}}+
c\sum\limits_{i=1}^{m}\theta_{i}\sum\limits_{j=1}^{m}\tilde{a}_{ij}
\frac{||\delta{y}_{j}(t)||_{1}}{||\delta{y}_{j}(t)||_{1}}
\\&
= L\kappa_{2}\max_{i}\{|\tilde{a}_{ii}|\}||\delta{y}(t)||_{\{\theta,1\}}
+c\sum\limits_{j=1}^{m}
\sum\limits_{i=1}^{m}\theta_{i}\tilde{a}_{ij}\\
&
< L\kappa_{2}\max_{i}\{|\tilde{a}_{ii}|\}||\delta{y}(t)||_{\{\theta,1\}}-cm\theta^{\star}
\end{align*}
In case that $c>\frac{L\kappa_{2}\max_{i}\{|\tilde{a}_{ii}|\}||\delta{y}(0)||_{\{\theta,1\}}}{m\theta^{\star}}$, we have
\begin{align*}
\frac{d
||\delta{y}(t)||_{\{\theta,1\}}}{dt}
&<L\kappa_{2}\max_{i}\{|\tilde{a}_{ii}|\}||\delta{y}(0)||_{\{\theta,1\}}-cm\theta^{\star} <0.
\end{align*}

Denote $\tilde{L}=L\kappa_{2}\max_{i}\{|\tilde{a}_{ii}|\}$, we have
$$||\delta{y}(t)||_{\{\theta,1\}}=0$$
when $t\ge \frac{||\delta{y}(0)||_{1}}{cm\theta^{\star}-\tilde{L}||\delta{y}(0)||_{\{\theta,1\}} }$.

{\bf Secondly}, we propose following continuous finite time convergence algorithm
\begin{eqnarray}
\left\{\begin{array}{ll}\frac{dx_1(t)}{dt}&=f(x_1(t))
+c\frac{\sum\limits_{j=1}^m\tilde{a}_{1j}(x_j(t)-s(t))}
{||\sum\limits_{j=1}^m\tilde{a}_{1j}(x_j(t)-s(t))||_{1}^{\alpha}},\\
\frac{dx_i(t)}{dt}&=f(x_i(t))+c\frac{\sum\limits_{j=1}^ma_{ij}x_j(t)}
{||\sum\limits_{j=1}^ma_{ij}x_j(t)||_{1}^{\alpha}},~~~~i=2,\cdots,m \end{array}\right.
\end{eqnarray}
where $0<\alpha<1$.

By similar derivations, we have
\begin{align*}
\frac{d
||\delta{y}(t)||_{\{\theta,1\}}}{dt}&\le \tilde{L}\sum_{i=1}^{m}\theta_{i}
||\delta y_{i}(t)||_{1}+c\sum\limits_{j=1}^{m}
\sum\limits_{i=1}^{m}\theta_{i}\tilde{a}_{ij}||\delta{y}_{j}(t)||_{1}^{1-\alpha}\\
& \le \tilde{L}\sum_{i=1}^{m}\theta_{i}
||\delta y_{i}(t)||_{1}-c\theta^{\star}\sum\limits_{j=1}^{m}
||\delta{y}_{j}(t)||_{1}^{1-\alpha}
\end{align*}

By Jensen inequality
\begin{align*}
(\sum\limits_{j=1}^{m}||\delta{y}_{j}(t)||_{1})^{1-\alpha}\le m^\alpha(\sum\limits_{j=1}^{m}||\delta{y}_{j}(t)||_{1}^{1-\alpha})
\end{align*}
we have
\begin{align*}
\frac{d
||\delta{y}(t)||_{\{\theta,1\}}}{dt}&<\tilde{L}\sum_{i=1}^{m}\theta_{j}
||\delta y_{i}(t)||_{1}-c\frac{\theta^{\star} }{m^\alpha}(\sum\limits_{j=1}^{m}\theta_{i}||\delta{y}_{j}(t)||_{1})^{1-\alpha}\\
&=\tilde{L}||\delta{y}(t)||_{\{\theta,1\}}-c\frac{\theta^{\star} }{m^\alpha}||\delta{y}(t)||_{\{\theta,1\}}^{1-\alpha}
\end{align*}

\begin{propo}(Lemma 2 in \cite{Finite18})~~Suppose the Lyapunov function $V(x)$ is defined on a neighborhood $U$ of the origin, and satisfies
\begin{align*}
\dot{V}(x)\le -\beta V^p(x)+kV(x), 0<p<1,
\end{align*}
then, in case that $\beta>kV(0)^{1-p}$, the origin is finite-time stable, and the settling time
\begin{align*}
\bar{t}=\int_{0}^{V(0)}\frac{dV}{\mu(V)}&=\int_{0}^{V(0)}\frac{dV}{\beta V^p(1-\frac{k}{\beta}V^{1-p})}\\
&=\frac{\ln(1-\frac{k}{\beta}V(0)^{1-p})}{k(p-1)}.
\end{align*}
\end{propo}
Let $V=||\delta{y}(t)||_{\{\theta,1\}}$, $k=\tilde{L}$, $p=1-\alpha$, $\beta=c\frac{\theta^{\star} }{m^\alpha}$, $c>\frac{\tilde{L}m^\alpha}{\theta^{\star}}||\delta{y}(0)||_{\{\theta,1\}}^{\alpha}$, the settling time
\begin{align*}
\bar{t}=
\frac{\ln(1-\frac{\tilde{L}m^\alpha}{c\theta^{\star}}||\delta{y}(0)||_{\{\theta,1\}}^{\alpha})}{-\alpha}
\end{align*}
and $\delta y(t)=0$ if $t>\tilde{t}$. Theorem is proved.

\section{Heterogeneous dynamic networks}

As an application of previous results, in this section, we discuss synchronization for heterogeneous dynamic networks.

Given $m+1$ systems $\dot{x}(t)=f_{k}(x(t))$, $k=0,1,\cdots,m$, where $f_{k}(x)$, $i=0,1,\cdots,m$, are (maybe different) bounded continuous functions.

Consider following system
\begin{eqnarray}\label{pinfl3b}
\left\{\begin{array}{ll}\frac{dx_{1}(t)}{dt}&=f_{1}(x_{1}(t))
+c_{02}\frac{\sum\limits_{j=1}^m\tilde{a}_{1j}(x_{j}(t)-s(t))}
{||\sum\limits_{j=1}^m\tilde{a}_{1j}(x_{j}(t)-s(t))||_{1}},\\
\frac{dx_{i}(t)}{dt}&=f_{i}(x_{i}(t))+c_{02}\frac{\sum\limits_{j=1}^m\tilde{a}_{ij}x_{j}(t)}
{||\sum\limits_{j=1}^m\tilde{a}_{ij}x_{j}(t)||_{1}},~~~~i=1,2,\cdots,m \end{array}\right.
\end{eqnarray}
where $k=1,\cdots,m,$ $s(t)$ is a solution satisfying $\dot{s}(t)=f_{0}(s(t))$.

\begin{theorem}~~Suppose $||f_{0}(x_i(t))-f_{0}(s(t))||_{1}\le L||x_i(t)-s(t)||_{1}$.
Algorithm (\ref{pinfl3b}) can synchronize all $x_{i}(t)$ to $s(t)$ in finite time.
\end{theorem}
{\bf Proof}~~Denote $y_{i}(t)=\sum\limits_{j=1}^{m}\tilde{a}_{ij} x_{j}(t)$, $\delta{y}_{i}(t)=\sum\limits_{j=1}^{m}\tilde{a}_{ij} (x_{j}(t)-s(t))$.
 Algorithm (\ref{pinfl3b}) can be rewritten as
\begin{align}\label{pinfl3c}
\dot{\delta}y_{i}(t)=&\sum\limits_{j=1}^{m}\tilde{a}_{ij}[f_{j}(x_{j}(t))-f_{0}(s(t))]
+c_{02}\sum\limits_{j=1}^{m}\tilde{a}_{ij} \frac{\delta{y}_{j}(t)}{||\delta{y}_{j}(t)||_{1}}\nonumber
\\=& \sum\limits_{j=1}^{m}\tilde{a}_{ij}[f_{0}(x_{j}(t))-f_{0}(s(t))]
+c_{02}\sum\limits_{j=1}^{m}\tilde{a}_{ij} \frac{\delta{y}_{j}(t)}{||\delta{y}_{j}(t)||_{1}}+u_{i}(t)
\end{align}
where $u_{i}(t)=\sum_{j=1}^{m}\tilde{a}_{ij}[f_{j}(x_{j}(t))-f_{0}(x_{j}(t))]$ and there is a constant $\tilde{c}>0$ such that $||u_{i}(t)||_{1}\le \tilde{c}$ for all $i=1,\cdots,m$.

Similar to the proof of Theorem 1, define
\begin{eqnarray*}
||\delta{y}(t)||_{\{\theta,1\}}=\sum_{i=1}^{m}\theta_{i}
||\delta{y}_{i}(t)||_{1}=\sum_{i=1}^{m}\theta_{i}
\sum_{k=1}^{n}|\delta{y}_{i}^{k}(t)|
\end{eqnarray*}
we have
\begin{align*}
&\frac{d
||\delta{y}(t)||_{\{\theta,1\}}}{dt}
< L\kappa_{2}\max_{i}\{|\tilde{a}_{ii}|\}||\delta{y}(t)||_{\{\theta,1\}}-c_{02}m\theta^{\star}+\tilde{c}m
\end{align*}
In case that $c_{02}>\frac{L\kappa_{2}\max_{i}\{|\tilde{a}_{ii}|\}||\delta{y}(0)||_{\{\theta,1\}}+\tilde{c}m}{m\theta^{\star}}$, we have
\begin{align*}
\frac{d
||\delta{y}(t)||_{\{\theta,1\}}}{dt}
&<L\kappa_{2}\max_{i}\{|\tilde{a}_{ii}|\}||\delta{y}(0)||_{\{\theta,1\}}-c_{02}m\theta^{\star}+\tilde{c}m <0
\end{align*}
Therefore, $||\delta{y}(t)||_{\{\theta,1\}}=0$, when $t>\frac{||\delta{y}(0)||_{\{\theta,1\}}}
{L\kappa_{2}\max_{i}\{|\tilde{a}_{ii}|\}||\delta{y}(0)||_{\{\theta,1\}}-c_{02}m\theta^{\star}+\tilde{c}m}$. %

\section{Conclusions}
In this paper, we propose new distributed pinning synchronization models of complex networks with a single nonlinear pinning controller. One is discontinuous algorithm. The other one is continuous algorithm. All two algorithms can synchronize all states of individual nodes to a specified trajectory in finite time. The result can also apply to heterogeneous dynamic networks. The coupling matrix is asymmetric (or the connecting graph is directed). The method is innovative. A coordinate transform is introduced. Different from Lyapunov function with $L_{2}$ norm, here, $L_{1}$ norm with weights and without weights are introduced, which is proved to be very effective.


\noindent{\bf References}
\begin{thebibliography}{99}
\bibitem{Chen1}
T. Chen, X. Liu, and W. Lu, Pinning complex networks by a single controller, \emph{IEEE Transactions on Circuits and Systems I-Regular Papers}, vol. 54, no. 6, pp. 1317-1326, Jun. 2007.

\bibitem{Chen2}
W. Lu and T. Chen, New approach to synchronization analysis of linearly coupled ordinary differential systems, \emph{Physica D}, vol. 213, no. 2, pp. 214-230, Jan. 2006.

\bibitem{Finite86}
V. T. Haimo, Finite time controllers, \emph{SIAM Journal on Control and Optimization}, vol. 24, no. 4, pp. 760-770, Jul. 1986.

\bibitem{Finite98}
S. Bhat and D. Bernstein, Continuous finite-time stabilization of the translational and rotational double integrators, \emph{IEEE Transactions on Automatic Control}, vol. 43, no. 5, pp. 678-682, May 1998.

\bibitem{Finite02}
Y. Hong, Y. Xu, and J. Huang, Finite-time control for robot manipulators, \emph{Systems $\&$ Control Letters}, vol. 46, no. 4, pp. 243-253, Jul. 2002.

\bibitem{Finite05}
W. Lu and T. Chen, Dynamical behaviors of Cohen-Grossberg neural networks with discontinuous activation functions, \emph{Neural Networks}, vol. 18, no. 3, pp. 231-242, Apr. 2005.

\bibitem{Finite12}
A. Polyakov, Nonlinear feedback design for fixed-time stabilization of linear control systems, \emph{IEEE Transactions on Automatic Control}, vol. 57, no. 8, pp. 2106-2110, Aug. 2012.

\bibitem{Finite15}
A. Polyakov, D. Efimov, and W. Perruquetti, Finite-time and fixed-time stabilization: Implicit Lyapunov function approach, \emph{Automatica}, vol. 51, pp. 332-340, Jan. 2015.

\bibitem{Finite16}
W. Lu, X. Liu, and T. Chen, A note on finite-time and fixed-time stability, \emph{Neural Networks}, vol. 81, pp. 11-15, Sep. 2016.

\bibitem{Finite06}
J. Cort$\acute{e}$s, Finite-time convergent gradient flows with applications
to network consensus, \emph{Automatica,} vol. 42, no. 11, pp. 1993-2000,
Nov. 2006.

\bibitem{Finite09}
F. Xiao, L. Wang, J. Chen, and Y. Gao, Finite-time formation control for multi-agent systems, \emph{Automatica}, vol. 45, no. 11, pp. 2605-2611, Nov. 2009.

\bibitem{Finite13}
X. Yang, Z. Wu, and J. Cao, Finite-time synchronization of complex networks with nonidentical discontinuous nodes, \emph{Nonlinear Dynamics,} vol. 73, no. 4, pp. 2313-2327, Sep. 2013.

\bibitem{Finite18}
X. Liu and T. Chen, Finite-time and fixed-time cluster synchronization with or without pinning control, \emph{IEEE Transactions on Cybernetics}, vol. 48, no. 1, pp. 240-252, Jan. 2018.

\bibitem{Finite19}
G. Wen, P. Wang, X. Yu, W. Yu, and J. Cao, Pinning synchronization of complex switching networks with a leader of nonzero control inputs, \emph{IEEE Transactions on Circuits and Systems I: Regular Papers}, vol. 66, no. 8, pp. 3100-3112, Aug. 2019.

\bibitem{Finite20}
S. Yang, C. Hu, J. Yu, and H. Jiang, Finite-time cluster synchronization in complex-variable networks with fractional-order and nonlinear coupling, \emph{Neural Networks,} vol. 135, pp. 212-224, Mar. 2021.

\bibitem{Finite20a}
L. Hua, S. Zhong, K. Shi, and X. Zhang, Further results on finite-time synchronization of delayed inertial memristive neural networks via a novel analysis method, \emph{Neural Networks,} vol. 127, pp. 47-57, Jul. 2020.

\bibitem{Finite21}
L. Hua, H. Zhu, K. Shi, S. Zhong, Y. Tang, and Y. Liu, Novel finite-time reliable control design for memristor-based inertial neural networks with mixed time-varying delays, \emph{IEEE Transactions on Circuits and Systems I-Regular Papers}, vol. 68, no. 4, pp. 1599-1609, Apr. 2021.


















\end{thebibliography}
\end{document}